# The Promise and Peril of Generative AI: Evidence from GPT as Sell-Side Analysts*

Edward Xuejun Li[a], Min Shen[b], Zhiyuan Tu[c], Dexin Zhou[d]

## Abstract

Large language models (LLMs) promise to democratize financial analysis by reducing information-processing costs. Yet equal access does not ensure equal outcomes, as the locus of friction may shift from *processing* information to *evaluating* model outputs. We study GPT's earnings forecasts following corporate earnings releases and document two patterns. First, GPT's narrative attention is consistent and human-like but not always associated with higher forecast accuracy. Second, its quantitative reasoning varies substantially across contexts, challenging the view that LLMs are uniformly weak at numerical tasks. Building on these insights, we propose a diagnostic framework that links forecast accuracy to observable processing features (i.e., narrative focus, numerical reasoning, and self-assessed confidence). These indicators serve as proxies for this new form of information friction and alert investors when to exercise caution. Our study has implications for information frictions, regulatory oversight, and the economics of AI-mediated financial markets.

---

* We are grateful for the feedback from Qiang Cheng, Stan Markov, Yue Zhao (discussant), workshop participants at Baruch College, Central South University, Chinese University of Hong Kong (Shenzhen), Florida International University, Hunan University, Sun Yat-Sen University, Southwestern University of Finance and Economics, Vanderbilt University, the AAA Annual Meeting, Hawaii Accounting Research Conference, and the 2024 Wolfe Research QES NLP and Machine Learning in Investment Management Conference.
[a] Stan Ross Department of Accountancy, Baruch College, Email: edward.li@baruch.cuny.edu
[b] Costello College of Business, George Mason University, Email: mshen3@gmu.edu
[c] School of Accounting, Southwestern University of Finance and Economics, Email: zhiyuantu@swufe.edu.cn
[d] Bert Wasserman Department of Economics and Finance, Baruch College, Email: dexin.zhou@baruch.cuny.edu

## 1 Introduction

In financial markets, persistent disparities in information-processing capacity continue to shape investor behavior and asset pricing. Theories attribute these frictions to differences in data access, analytical skill, or processing costs (Grossman and Stiglitz, 1980; Verrecchia, 2001; Blankespoor, deHaan, and Marinovic, 2020). The advent of large language models (LLMs), such as OpenAI's GPT, may change these dynamics by making advanced analytical tools broadly accessible at low marginal costs. Recent studies suggest that LLMs reduce information processing costs for both professional and non-professional users (e.g., Bertomeu, Lin, Liu, and Ni, 2025; Bradshaw, Ma, Yost, and Zou, 2025; Cheng, Lin, and Zhao, 2025; deHaan, Lee, Liu, and Noh, 2025; Sheng, Sun, Yang, and Zhang, 2025). This equalized accessibility has led to optimism that LLMs could democratize financial analysis, allowing non-experts to engage with sophisticated analytical tasks (Chang, Dong, Martin, and Zhou, 2025).

Equal access, however, does not ensure equal outcomes. While LLMs reduce traditional barriers to information processing, they may introduce a new friction: the difficulty of assessing output quality. Because LLMs often provide responses without transparent reasoning, users, especially non-experts, could struggle to identify reliable outputs.[1] Recent evidence illustrates this challenge. In 2024, the Securities and Exchange Commission (SEC) sanctioned two advisory firms for misleading claims about AI-powered forecasts.[2] Although neither firm possessed functioning AI models, investors acted on their recommendations, suggesting limited ability to evaluate these outputs. Experimental evidence by Klingbeil, Grützner, and Schreck (2024) similarly shows that investors often over-rely on AI advice even when it contradicts human judgment, leading to inefficient investment decisions. When investors cannot assess model performance, this new friction may counteract LLMs' potential to reduce information disparities, instead amplifying them among less sophisticated users (Barber, Odean, and Zhu, 2009; Zhu, He, and Hu, 2023).

---

[1] A 2020 report from the Financial Industry Regulatory Authority (FINRA) titled "Artificial Intelligence (AI) in the Securities Industry" notes that "[s]ome ML models, however, are described as 'black boxes' … because it may be difficult or impossible to explain how the model works (i.e., how its predictions or outcome are generated)." https://www.finra.org/sites/default/files/2020-06/ai-report-061020.pdf, accessed on October 10, 2025.

[2] SEC Settles With Two Investment Advisers Over Alleged 'AI Washing'—The regulator said Delphia and Global Predictions made 'false and misleading claims' about the use of AI in their investment platforms, *Wall Street Journal*, March 18, 2024. https://www.wsj.com/articles/sec-settles-with-two-investment-advisers-over-alleged-ai-washing-cd067049?utm_source=chatgpt.com, accessed on August 23, 2025.

These observations motivate our research questions: under what conditions do LLMs produce higher-quality outputs, and how can observable signals of their information processing be linked to output quality? More broadly, we ask whether these signals can be organized into a diagnostic framework that alerts users when to use caution, thereby suggesting contexts in which LLM-based analysis may amplify, rather than mitigate, information frictions.

We study this question by evaluating GPT, the most widely used LLM,[3] on a realistic and economically relevant task: forecasting one-year-ahead earnings based on firms' earnings press releases. Our design approximates how non-experts interact with GPT, allowing us to assess its potential to democratize financial analysis. First, we employ simple, non-technical prompts in natural language to evaluate GPT's performance in a setting where domain knowledge is limited. Second, we examine how GPT processes information—a critical yet understudied aspect given the model's proprietary nature. Specifically, we focus on how it analyzes two distinct disclosure components (i.e., narrative content and tabular financial statements) and assess how each relates to forecast outcomes. Third, we compare GPT's forecasts to contemporaneous analyst forecasts, which provide a natural benchmark and help control for unobserved firm- and macro-level factors. Finally, to mitigate look-ahead bias and better reflect users' real-time decision environment, we restrict our sample to press releases near GPT's knowledge cutoff.[4]

[3] A recent survey by Rainie (2025) indicates that GPT remains the most widely used model among LLM users.
[4] Interpreting pre-cutoff performance requires caution. Two factors may give the model what we call a "time-traveler's advantage." First, historical disclosures may contain subtle cues, such as phrasing, structure, or context, that inadvertently signal future outcomes. Second, GPT may draw on analytical approaches and domain knowledge that were developed only after the disclosure period, allowing it to evaluate historical information in ways not available to contemporaneous analysts. Recognizing these effects is important for a fair assessment of GPT's real-time forecasting capability.

Our analysis draws on 7,114 earnings press releases issued by 1,000 randomly selected firms over a two-year window centered on GPT-4 Turbo's knowledge cutoff date of April 30, 2023. This sampling strategy balances access to the latest model with sufficient post-cutoff data to measure realized forecast accuracy.[5] For each press release, we prompt GPT to (1) identify the three sentences most relevant for forecasting, (2) analyze financial statements and identify three key metrics (ratios or trends) from its quantitative analysis, and (3) predict one-year-ahead GAAP earnings per share (EPS). These tasks allow us to study GPT's narrative attention, quantitative reasoning, and predictive accuracy in a structured and replicable setting. We aim to diagnose systematic patterns in GPT's information processing that correlate with variation in its forecasting performance, rather than to establish causal relationships.

We first document a clear distinction in how GPT processes narrative versus quantitative information. Narrative attention is highly consistent across firms: GPT tends to prioritize sentences that are placed early in the text, readable, forward-looking, and negatively toned—attributes tied to investor attention and value relevance (Rawson, Twedt, and Watkins, 2024; Li, 2008; Elliott, Rennekamp, and White, 2015; Loughran and McDonald, 2011).[6] This consistency likely reflects GPT's training on large-scale textual corpora, which embed strong priors about linguistic salience. In contrast, quantitative reasoning varies widely across firms, reflecting heterogeneity in domain-specific learning. GPT's ability to approximate expert judgment (i.e., identifying the key metrics that analysts tend to emphasize) is higher for firms with more reliable quantitative disclosures and proxies for stronger exposure during training (e.g., broader analyst coverage).[7] These results suggest that GPT capabilities are context-dependent: stable in narrative processing but more variable and contingent in quantitative analysis. The latter finding challenges the view that LLMs struggle broadly at numerical reasoning in financial context.[8]

[5] Our primary analysis relies on GPT-4 Turbo, whose knowledge cutoff allows us to construct a complete post-cutoff fiscal year (i.e., four consecutive quarters) for which realized one-year-ahead forecast errors can be calculated as of July 2025. To ensure that our inferences are not model-specific, we replicate the analysis with GPT-4o. Despite the shorter post-cutoff window and greater potential for look-ahead bias, the main inferences remain unchanged (see Section 4.5.2).

[6] Interestingly, GPT also favors longer sentences, diverging from the typical human preference for brevity (Whelan, 2020). This likely reflects its capacity to process complex linguistic structures.

[7] Liu et al. (2024) evaluate LLMs' statistical and causal reasoning using real-world data, finding a 58 percent accuracy rate for models like GPT-4. Their analysis relies heavily on structured, domain-specific data, with performance deteriorating in quantitative tasks when such data are sparse.

In our main analysis, we document novel links between GPT's information-processing patterns and its forecasting accuracy. Forecasts tend to be more accurate when GPT selects more readable and forward-looking sentences but less accurate when it favors negatively toned ones.[9] Thus, GPT's narrative attention is not always associated with higher accuracy, suggesting its mixed capacity to extract relevant cues from narratives. Forecast accuracy is also higher when GPT-identified metrics align with those emphasized by analysts: one standard deviation higher in alignment is associated with a 16.8 percent lower forecast error. These results suggest that variation in GPT's accuracy is economically significant and systematically correlated with observable aspects of its processing. When benchmarked against analysts, its accuracy is closer to analyst levels when metric alignment is high but falls short otherwise. Taken together, these findings move the discussion beyond average-case performance and toward a conditional view of when LLMs perform relatively well in financial tasks.

We also examine GPT's confidence scores, which capture the model's self-assessed certainty in its forecasts. Confidence is positively correlated with the quality of GPT's narrative and quantitative reasoning, suggesting that these scores carry informational value. However, while GPT's forecasting performance worsens after the knowledge cutoff, its confidence in those forecasts does not systematically decline, suggesting limited self-awareness of training constraints imposed by outdated information.

Building on these findings, we propose a diagnostic framework that maps GPT's observable processing features into a composite reliability score. Each feature (i.e., selecting readable, forward-looking, or positively toned narratives; identifying analyst-consistent financial metrics; and reporting high confidence) receives a binary indicator for whether GPT's behavior aligns with characteristics of higher-quality analysis. The resulting composite score increases monotonically with forecast accuracy. This framework offers an interpretable, scalable approach to assessing when GPT's forecasts are more or less accurate and provides proxies to guide investors, especially non-experts.

---

[8] For example, Levy (2025) states that "I systematically examine the performance of LLMs on numerical reasoning tasks relevant to accounting and finance. I show that LLMs perform poorly in this domain and therefore should be applied judiciously." (p. 7)

[9] Crowley and Wong (2022) show that negative tone, while salient, often introduces considerable noise into predictions. This suggests that GPT can overweight salient cues at the cost of accuracy.

We conduct a series of supplementary analyses to assess the stability and scope of our inferences. First, our results generalize to non-GAAP earnings targets and a shorter time horizon (i.e., one quarter ahead), both in absolute terms and relative to analysts. Second, re-estimating our main tests with the more recent GPT-4o model produces consistent results, despite a shorter post-cutoff period and greater scope for look-ahead bias. Finally, our inferences remain unchanged when using more sophisticated prompts modeled on the Chartered Financial Analyst (CFA) curriculum rather than simple instructions. Collectively, these analyses suggest that our findings are not driven by test specification, model version, or prompt sophistication.

Our results contribute to several strands of research in accounting and information economics. First, they speak to classic theories of information frictions (Grossman and Stiglitz, 1980; Sims, 2003; Verrecchia, 2001). While LLMs lower traditional processing costs, they may introduce a new friction: the need to assess the quality of model outputs. Because performance varies systematically across contexts, the binding constraint in LLM-mediated markets may shift from information processing to output assessment. Investors who cannot recognize when model reliability weakens may be disadvantaged if they rely on its output. Our proposed diagnostic signals can serve as proxies for the cost of output assessment, empirically capturing this new form of information friction.

Second, we add to the emerging literature on the democratization of financial analysis through generative AI. Blankespoor, Croom, and Grant (2024) document that while retail investors are increasingly adopting AI tools, the benefits they derive differ with user sophistication. Our finding of systematic variation in GPT's performance—linked to how it processes narrative and quantitative inputs—sheds light on where these disparities are likely to arise. We also contribute to the debate on AI's effects on investor information processing. Cheng et al. (2025) find narrowed bid-ask spreads during GPT outages, consistent with AI facilitating informed trading. In contrast, Chang et al. (2025) show that AI helps retail investors trade more on fundamentals, narrowing information gaps. Our results suggest that the net effect of AI likely varies with the trade-off between reduced processing costs and increased demands on abilities to evaluate outputs.

Third, our results also relate to the literature on delegation and monitoring in capital markets (Diamond, 1984; Holmström and Milgrom, 1991). While prior work focuses on human intermediaries such as analysts, credit rating agencies, or the financial press, our results shed light on delegation to algorithmic agents. GPT's effectiveness, like that of human delegates, varies systematically with the domain and quality of information inputs, suggesting domain-specific risks in algorithmic delegation.

Finally, our findings have implications for corporate disclosure and regulatory oversight. GPT's accuracy is higher when firms have broader analyst coverage and more reliable disclosures, conditions that provide rich learning signals and align with expert judgment. These patterns suggest two regulatory priorities: (1) improving disclosure quality and (2) strengthening investor protection, both by fostering model transparency (e.g., explaining data sources, training limits, and usage rules) and by building investors' ability to understand and use models wisely. Our proposed diagnostic framework represents an initial step, offering a structured way for users to assess the quality of model outputs. By prompting the model to articulate features of its own processing, users can obtain valuable signals correlated with output quality. These signals are not a substitute for expert judgment but can help identify when model-generated information warrants extra caution. Although our analysis focuses on earnings forecasts, the conceptual logic of our framework may extend to other domains where models interpret both narrative and quantitative disclosures.

The remainder of the paper proceeds as follows. Section 2 describes the sample and prompt design. Section 3 examines GPT's processing of narrative and tabular disclosures. Section 4 presents results of GPT's forecasts, and Section 5 proposes a diagnostic framework. Section 6 concludes.

## 2 Data and Prompt Design

### *2.1 Data*

Our initial sample comprises 68,514 quarterly earnings announcements from firms in the intersection of Compustat, CRSP, and I/B/E/S, spanning four quarters before and four quarters after GPT-4 Turbo's knowledge cutoff date of April 30, 2023. We obtain the full text of earnings press releases from 8-K filings on EDGAR, restricting attention to Item 2.02 (earnings press releases). This focus on full-text disclosures reduces the sample to 49,192 observations.

To balance computational feasibility and representativeness, we randomly select 1,000 firms with at least one quarter of data both before and after the cutoff date. We exclude firm-quarters without I/B/E/S GAAP earnings forecasts (GPS) and arrive at 7,216 observations with the necessary test variables.[10] GPT successfully produces GAAP EPS forecasts for 7,114 of these cases. The associated earnings press releases contain 329,333 sentences, providing large narrative data for our analysis.

[10] We use analysts' GAAP EPS forecasts in our analyses because GPT's EPS predictions are GAAP-based. This alignment avoids comparing GPT's GAAP forecasts to I/B/E/S street earnings, which adjust for non-GAAP items. As Bradshaw et al. (2018) emphasize, this distinction is critical, and nearly all analysts now provide GAAP EPS forecasts, making them an appropriate benchmark. In a robustness test, we also prompt GPT to produce non-GAAP earnings forecasts and compare them to I/B/E/S street earnings, obtaining similar results.

### *2.2 Prompt Design*

We use prompts that approximate how non-experts would typically interact with a language model: simple, conversational, and free of technical jargon—essentially providing relevant information and asking direct questions.

The prompts unfold in three stages. First, we introduce the overall task, specifying that GPT will analyze two types of inputs: narrative content from an earnings press release and tabular data from the accompanying financial statements. This establishes the scope and structure of the assignment. Second, we provide the cleaned narrative, extracted from the start of the release through the boilerplate forward-looking disclaimer section, with HTML tags and unstandardized financial tables removed. This helps restrict GPT's narrative processing to the substantive, human-readable disclosure. Third, we supply balance sheet and income statement information, following a practice similar to recent studies (e.g., Kim, Muhn, and Nikolaev, 2024). Specifically, GPT receives: (i) annual data—current and prior year for the balance sheet, and current plus two prior years for the income statement; and (ii) quarterly data[11]—current and prior quarter for the balance sheet, and current plus two prior quarters for the income statement, for a total of 10 financial statements.

The model is then instructed to perform three related tasks:

(1) Narrative selection – Identify the three most informative sentences for predicting future earnings and briefly justify each choice.
(2) Quantitative reasoning – Conduct quantitative analysis using the tabular financial data and identify the three metrics (financial ratios or trends other than EPS) produced that are relevant to the task, showing formulas and calculations.

[11] Although our forecasts target annual earnings, we also provide quarterly statements to capture timely current-quarter information.

(3) Forecasting – Predict both the direction and the numerical value of one-year-ahead GAAP and non-GAAP EPS, along with a confidence score for each forecast.

These tasks necessarily operate at different levels. Narrative salience is a sentence-level construct, as readers attend to cues such as readability and forward-looking orientation embedded in individual sentences. Quantitative reasoning, by contrast, is a firm–period activity that synthesizes ratios and trends across financial statements for a given quarter. Imposing a uniform unit could misrepresent at least one task: ratios are not sentence-level objects, and sentence-level salience is not captured by firm-level aggregates. Accordingly, we evaluate narrative selection at the firm–quarter–sentence level and quantitative reasoning at the firm–quarter level.

Spot-checks of GPT's outputs suggest that the model follows instructions, generating coherent reasoning chains and relevant selections. For transparency and replicability, we provide the full prompts and sample responses in Internet Appendix Tables IA1–IA2. While our prompts are informal and less sophisticated relative to the guide-by-example, chain-of-thought (CoT) prompting, they approximate natural non-expert usage. Prior work shows that while rigid demonstrations suit highly structured tasks such as arithmetic operations, minimal cues are often sufficient to elicit step-by-step reasoning, whereas lengthy exemplars may introduce distracting context.[12] We therefore treat simple prompting as the baseline and assess robustness with CFA-inspired prompts in a supplementary test.

## 3 GPT's Information Processing

To investigate how GPT processes information, we conduct two complementary analyses at distinct levels. For narrative disclosures, we examine whether GPT's selected sentences from the earnings press release align with or diverge from human attention. For quantitative disclosures, we analyze how GPT processes, focusing on whether the metrics it emphasizes correspond to those highlighted by domain experts (i.e., sell-side analysts).

[12] See, e.g., guide-by-example CoT (Wei et al., 2023), "let's think step by step" prompting (Kojima et al., 2022), and evidence that excessive context can reduce accuracy (Liu et al., 2024; Kandpal et al., 2023).

### *3.1 Narrative Attention*

A central component of GPT's forecasting process is to determine which parts of an earnings press release it should prioritize. These prioritization patterns are shaped by the model's large-scale pretraining and fine-tuning and may mirror—but not exactly replicate—those observed in human financial analysis. Understanding this process serves two purposes in our setting. First, it reveals which narrative attributes GPT tends to prioritize when distilling lengthy disclosures into focal inputs. Second, it allows us to evaluate whether these attributes are associated with variation in GPT's forecasting performance.

We focus on the three sentences in each press release that GPT identifies as most relevant for forecasting and examine the determinants of its selection. Specifically, we estimate a sentence-level linear probability model (LPM) as follows:

$$Picked_By_GPT_{i,t,s} = \beta_1 Order_{i,t,s} + \beta_2 Log(Length)_{i,t,s} + \beta_3 Fog_{i,t,s} + \beta_4 Guidance_{i,t,s} + \beta_5 Sentiment_{i,t,s} + Press\ Release\ Fixed\ Effects + \varepsilon_{i,t,s} \quad (1)$$

where the dependent variable, *Picked_By_GPT*, equals one if the sentence *s* in firm *i*'s earnings press release for quarter *t* is among the top three identified by GPT for forecasting. The explanatory variables are drawn from prior research on disclosure and capital markets. Order of appearance (*Order*) captures the primacy effect documented in investor reading patterns, where information presented earlier receives disproportionate attention (Rawson et al., 2024). Sentence length, measured as the log of word count (*Length*), may signal greater informational content but can also increase cognitive load (SEC, 1998; Whelan, 2020). Readability (*Fog* index) proxies for language clarity, which has been linked to both investor engagement and disclosure informativeness (Li, 2008).

Forward-looking statements and tone are also important attributes. The presence of management guidance (*Guidance*) may draw attention by shaping expectations and reducing uncertainty (Elliott et al., 2015). *Sentiment*, measured following Loughran and McDonald (2011), captures tone, with more negative language shown to provide incremental predictive content for future performance (e.g., Tetlock, 2007). Together, these attributes reflect well-established determinants of investor attention and value relevance. Testing their relation to GPT's sentence selections allows us to assess whether the model's narrative focus parallels that of human readers in financial analysis, and to identify systematic departures that may be predictive of variation in GPT's forecasting performance. Appendix A defines all variables. Press release fixed effects absorb firm- and time-invariant features of the press releases, eliminating the need for additional firm- or time-level controls.

Panel A of Table 1 reports summary statistics for the 329,333 sentences in our sample. On average, GPT selects 5.8 percent of sentences from a given press release. A typical sentence contains 31.5 words, exhibits a slightly positive tone (mean sentiment score of 0.003), and references guidance-related content in 5.4 percent of cases. Sentence order is evenly distributed by construction, while both sentence length and Fog index show substantial variation, providing ample dispersion for testing selection patterns.

Panel B presents the sentence-level regression results from estimating model (1). Results are reported for the full sample and for subsamples partitioned by the median values of firm size and analyst following, as well as across pre- and post-knowledge cutoff periods. These subsample analyses assess whether selection patterns vary with firms' information environment or across time.

Column (1) reports results on the full sample. We find that sentence order is negatively associated with selection: moving from the start to the end of a press release lowers the probability of selection by about 11.0 percent (t = -40.73). Readability also matters, with a one-standard-deviation increase in the Fog index decreasing the selection probability by 3.44 percent (-0.005 × 6.889). Guidance-related sentences are strongly associated with a higher probability of selection (18.2 percent, t = 29.97). Tone is statistically significant (t = -5.70) but economically negligible: a one–standard-deviation decrease in sentiment increases the probability of selection by only 0.34 percent (-0.096 × 0.035). These preferences closely mirror those documented for human investors, consistent with prior evidence on investor attention and the value relevance of disclosures (Rawson et al., 2024; Li, 2008; Elliott et al., 2015; Loughran and McDonald, 2011). A notable departure is GPT's preference for longer sentences: a one-standard-deviation increase in log-transformed sentence length (0.588, untabulated) raises the selection probability by 5.2 percent (0.088 × 0.588), which differs from the typical human preference for brevity. This contrast may reflect GPT's ability to handle more complex linguistic structures.

We find that GPT's narrative preferences are highly consistent across subsamples. Coefficients on order, length, readability, guidance, and sentiment retain their expected signs and statistical significance, with only modest variation in magnitude. The negative association with sentiment is more pronounced for smaller firms, suggesting that GPT places greater weight on cautious language in information-scarce environments. Overall, the stability of these results—across both the full sample and subsamples—suggests that GPT's narrative attention is systematic. In many respects, it resembles human preferences for early, readable, forward-looking, and negatively toned information, while uniquely favoring longer, information-dense sentences.

### *3.2 Quantitative Reasoning*

A second dimension of GPT's information processing is its engagement with structured, tabular financial data. In our prompt design, GPT reviews standardized balance sheet and income statement data after analyzing the narrative disclosures, and is then instructed to perform a quantitative analysis and report three non-EPS financial metrics (e.g., ratios or trends) it deems most relevant for forecasting. These metrics are produced in the course of forming the forecast. While they may not constitute intermediate "reasoning steps" in a strict sense, they nevertheless offer a useful window into the model's numerical reasoning patterns.

This analysis is motivated by the observation that, despite the view that LLMs struggle broadly with quantitative reasoning in financial contexts, little is known about the conditions under which such reasoning performs well or poorly. By examining the financial metrics GPT identifies, we can examine whether its quantitative selections resemble measures that professional analysts emphasize. This approach moves beyond testing whether GPT can "do the math" to evaluating its ability to identify which metrics warrant attention—a core aspect of expert financial judgment.

To benchmark GPT's metric selections, we compare them with the top three non-EPS metrics forecasted by sell-side analysts recorded in the I/B/E/S Detail database for the same firm–quarter. Because analysts' metric choices reflect accumulated domain expertise and established practice, the degree of overlap provides a proxy for GPT's ability to exert expert judgment. We capture this alignment with *MetricOverlap*, defined as the number of overlapping metrics between GPT and analysts, ranging from zero to three.

Panel A of Internet Appendix Table IA3 reports that GPT identifies "top three" metrics across 7,114 firm-quarters, which cluster into a manageable set of categories. Among these ratios and trends, the most frequently identified metric by GPT is net profit margin (71.9%), followed by return on equity (68.4%) and leverage ratio (33.8%). As Panel B illustrates, analysts, however, primarily emphasize revenue (66.2%), EBIT (61.1%), and net income (56.1%). Despite the differences, overlap is substantial: Panel C shows that in 69.6 percent of firm-quarters, GPT and analysts share at least one of their top three metrics.

To investigate the factors influencing GPT's alignment with analysts in quantitative analysis, we estimate the following Ordinary Least Squares (OLS) regression at the firm-quarter level for firm *i* in quarter *t*:

$$
\begin{aligned}
MetricOverlap_{i,t} = \ & \beta_1 Log(FS_LineItems)_{i,t} + \beta_2 Benf_Law_Deviation_{i,t} \\
& + \beta_3 Log(Firm_Size)_{i,t} + \beta_4 Log(Analyst)_{i,t} + \beta_5 Post_{i,t} \\
& + Industry/Fiscal\ Quarter\ Fixed\ Effects + \varepsilon_{i,t}
\end{aligned} \tag{2}
$$

where the dependent variable *MetricOverlap* is defined above. Drawing on prior research, we include two sets of explanatory variables. The first concerns disclosure clarity and quality. We measure financial statement disaggregation using the number of non-zero line items (*FS_LineItems*) in the standardized balance sheet and income statement provided to GPT. Prior research links more disaggregated reporting to greater decision usefulness (D'Souza, Ramesh, and Shen, 2010; Chen, Miao, and Shevlin, 2015). We also measure numerical consistency using *Benf_Law_Deviation*, calculated as the absolute difference between the distribution of first digits in reported numbers and the theoretical Benford distribution (Nigrini, 2012). Smaller deviations indicate more internally consistent reporting, which may facilitate quantitative reasoning. We examine whether alignment is higher when disclosures are more detailed and numerically consistent.

The second set captures learning opportunities for GPT. Larger firms (*Firm_Size*) generally issue more extensive and widely disseminated public disclosures, increasing the likelihood that GPT encounters relevant firm information during pretraining. Similarly, greater analyst coverage (*Analysts*) generates additional interpretations, summaries, and forecasts in the public domain, further expanding GPT's exposure to expert financial reasoning. Both factors increase the probability that GPT absorbed domain-specific material during pretraining, potentially strengthening its quantitative reasoning in those contexts. We therefore expect alignment to increase with firm size and analyst coverage. Finally, we include an indicator *Post* for the post-cutoff period, as well as fiscal quarter and industry fixed effects to control for seasonality and sector-specific heterogeneity.[13]

Table 2 reports the summary statistics and regression results for GPT–analyst alignment in metric selection. As shown in Panel A, the median value of *MetricOverlap* is 2, indicating that in more than half of firm-quarters, GPT's top three metrics overlap with those of analysts in at least two cases. This relatively high baseline suggests that GPT's quantitative focus frequently aligns with professional judgment, even though prior literature has questioned LLMs' numerical reasoning capabilities. Firms in our sample report, on average, 182 non-zero financial statement line items across the 10 financial statements provided, reflecting substantial disclosure detail, while the mean Benford's law deviation is 0.367, leaving meaningful variation in numerical consistency across firms.

Panel B shows that alignment is systematically correlated with regressors. In Column (1), the coefficient on *Log(FS_LineItems*) is positive and highly significant (2.111, t = 7.11). Given a standard deviation of 0.091 in log-transformed line items (untabulated), this corresponds to an increase in *MetricOverlap* of about 0.192, or roughly 10 percent relative to the sample mean of 1.961. By contrast, the coefficient on *Benf_Law_Deviation* is negative and significant (-0.983, t = -4.92). A one-standard-deviation reduction in Benford's law deviation (0.096) is associated with a 4.8 percent decrease in *MetricOverlap* relative to the sample mean ($-0.983 \times 0.096/1.961$).

The estimated effects are stronger for firm size and analyst coverage. The coefficient on *Log(Firm_Size)* (0.156, t = 8.53), combined with a standard deviation of 1.291, implies that a one-standard-deviation increase in firm size is associated with a 10 percent increase in *MetricOverlap* relative to the sample mean. For *Log(Analyst)* (0.076, t = 2.08), the corresponding effect is a 2.3 percent increase. Both associations are consistent with the idea that larger firms and those with broader analyst coverage may have generated more public-domain material, which GPT could plausibly have encountered during pretraining. In Column (2), we re-estimate the model including a post-knowledge cutoff indicator (*Post*). Results remain unchanged, and the coefficient on *Post* is insignificant, suggesting no structural shift in alignment over time.

[13] Our focus is on whether GPT's quantitative reasoning differs with firm size, analyst coverage, and knowledge cutoff, not on how its determinants vary with these measures. Therefore, we do not estimate the model for subsamples partitioned on these measures, as in Table 1.

Taken together, these results highlight distinct patterns in GPT's information processing. Narrative attention appears stable across firms and over time, whereas the quality of quantitative reasoning is more context-dependent. GPT's ability to approximate analysts' judgment is higher with richer and more reliable information, consistent with the notion that its quantitative reasoning varies with both the clarity of the inputs and the extent of domain-relevant materials available during pretraining.

A potential concern is that GPT-identified top sentences and financial metrics may not reflect the cues it actually relies on when generating forecasts. To validate this, we prompt GPT to justify its top sentence and metric selections. For individual sentences, GPT must explain the specific cues it finds informative. For individual metrics, GPT must both justify their relevance narratively and display intermediate calculation steps.

We compile these justifications into word clouds, shown in Figure 1: Panel A for narrative analysis and Panel B for quantitative reasoning. The most prominent terms, such as revenue, growth, margin, earnings, and outlook, closely track established forecasting determinants in the accounting literature. This semantic alignment between GPT's stated rationale and its reported selections supports our interpretation that the processing measures capture meaningful aspects of the model's reasoning rather than purely post hoc rationalization. We next examine whether GPT's narrative attention and quantitative reasoning correlate with the systematic variation in its forecasting performance.

## 4 GPT's Forecasting Performance

Our primary objective is to evaluate GPT's own earnings forecasts—assessing the extent of cross-firm variation and testing whether this variation is systematically linked to its observable processing patterns. As a supplementary test, we also benchmark GPT's forecasts against those of sell-side analysts for the same earnings releases to mitigate concerns about correlated factors, such as firm fundamentals, disclosure attributes, or macroeconomic conditions, that may influence both GPT and analyst forecasts.

### *4.1 Univariate Analysis of GPT's Forecast Accuracy*

Our primary analysis focuses on GPT's GAAP EPS forecasts. For each press release, we calculate the absolute forecast error ($|FE_GPT|$) as:

$$|FE_GPT|_{i,t} = \frac{|Forecast_{i,t} - Actual_{i,t}|}{Price_{i,t-1}},$$

where $Forecast_{i,t}$ is GPT's one-year-ahead GAAP EPS forecast for firm *i* based on the earnings press release on day *t*, $Actual_{i,t}$ is the corresponding realized EPS obtained from Compustat,[14] and $Price_{i,t-1}$ is the firm's closing stock price on the day prior to the earnings release date. For comparison, we measure analysts' absolute forecast error $|FE_Analyst|_{i,t}$ using consensus GAAP EPS analyst forecast (GPS) from the I/B/E/S unadjusted summary database for the same fiscal period after the earnings announcement. To mitigate the influence of outliers, we winsorize both forecast accuracy variables at the 1st and 99th percentiles.

Panel A of Table 3 shows that GPT's forecast errors are widely dispersed across the 7,114 press releases in our sample. The standard deviation (0.141) is more than 5.6 times its median value (0.025), and the interquartile range is economically large. This variation is not driven by the model's knowledge cutoff: summary statistics of GPT's forecast errors before and after April 30, 2023 show similar dispersion. Untabulated variance tests show that the spread in forecast errors remains statistically indistinguishable across the pre- and post-cutoff subsamples, suggesting that variation in GPT's forecasting performance is a persistent feature rather than an artifact of knowledge staleness.

[14] We also draw actual GAAP EPS (GPS) from I/B/E/S and obtain consistent inferences.

Next, we examine whether this variation is systematically related to differences in GPT's information processing. Panel B of Table 3 shows that the forecast error |*FE_GPT*| declines monotonically with *MetricOverlap*. The mean |*FE_GPT*| decreases from 0.117 when overlap is zero to 0.061 at the highest overlap level, approximately 48 percent lower. When we further partition the sample by firm size (large vs. small) and analyst following (high vs. low), a similar pattern largely persists within each subgroup. This consistency suggests that the negative correlation between *MetricOverlap* and forecast error is not driven by firm characteristics but reflects a systematic link between GPT's metric selection and its forecasting accuracy.

We also benchmark GPT's forecast errors against those of professional analysts, using the difference |*FE_GPT*|-|*FE_Analyst*|. Panel C of Table 3 shows that the mean difference is positive, indicating that GPT underperforms analysts on average. However, the 25th percentile of this distribution is -0.001, implying that GPT outperforms analysts in more than one-fourth of firm-quarters in our sample. The mean difference is slightly smaller pre-cutoff (0.033) than post-cutoff (0.038), suggesting a modest deterioration in GPT's relative performance after the knowledge cutoff.

Finally, Panel D of Table 3 reports the mean forecast error difference across *MetricOverlap* levels and subsamples partitioned by firm size and analyst following. Consistent with the results in Panel B, GPT's relative performance improves as *MetricOverlap* increases. The mean forecast error difference between GPT and analysts declines from 0.063 at zero overlap to 0.028 at three, with the monotonic pattern strongest for larger firms and those with higher analyst following. These results collectively highlight the association between GPT's information processing, contextual richness, and variation in its forecasting performance.

*4.2 Multivariate Analysis of GPT's Forecast Accuracy*

We next estimate regressions of GPT's forecast accuracy on features of its narrative and quantitative information processing:

$$
\begin{aligned}
|FE_GPT_{i,t}| = & \sum \alpha \cdot GPT_Increm_Char_{i,t} + \beta \cdot MetricOverlap_{i,t} \\
& + \sum \gamma \cdot PR_General_Char_{i,t} + \theta \cdot Post_t + \lambda \cdot Log(Firm_Size)_{i,t} \\
& + \mu \cdot Log(Analyst)_{i,t} + Industry\ Fixed\ Effects \\
& + Fiscal\ Quarter\ Fixed\ Effects + \varepsilon_{i,t} \qquad (3).
\end{aligned}
$$

The first set of processing features captures the incremental characteristics of GPT's top three selected sentences (*GPT_Increm_Char*). By measuring differences relative to the average sentence in the same press release, we isolate GPT's unique narrative prioritization—whether it emphasizes earlier placement, longer or more readable sentences, forward-looking statements, or negative sentiment—and test whether these emphases are associated with its forecasting performance. The second processing feature, *MetricOverlap*, is defined as in the previous section. Higher alignment scores indicate GPT's greater ability to approximate expert judgment by focusing on financially salient metrics that historically carry predictive power for earnings.

We include *PR_General_Char* to control for the overall characteristics of the press release, *Post* to account for the potential change in performance as information grows stale after the knowledge cutoff. In addition, *Firm_Size* and *Analyst* capture differences in the richness of the underlying information environment and GPT's likely prior exposure to similar firms or disclosures, which may influence its learning opportunities beyond what *MetricOverlap* reflects. Industry and fiscal-quarter fixed effects further control for sectoral and seasonal heterogeneity. All variables are defined in Appendix A.

Table 4 Panel A provides summary statistics for these variables. GPT-selected sentences largely differ from the average sentence in the same press release, confirming the patterns noted in Table 1. On average, GPT-prioritized sentences appear 17.1 percent earlier in the press

release, are 38 words longer (before the logarithm transformation), have a Fog index 2.15 points lower (indicating greater readability), contain 18.1 percent more guidance-related content, and have a slightly more negative tone. Given the symmetric pre- and post-cutoff sample design, the *Post* variable has a mean close to 0.5.

Panel B presents the regression estimates. Column (1), which includes the main processing feature variables and the *Post* indicator, shows that both narrative attention and quantitative alignment are strongly associated with forecast accuracy. Among narrative features, GPT's focus on readability and forward-looking statements is positively associated with forecast accuracy. Specifically, a one-standard-deviation decrease in *GPT_Increm_Fog* (i.e., selecting more readable sentences) is associated with a 0.009 reduction in forecast errors (-4.302 × 0.002), while a one-standard-deviation increase in *GPT_Increm_Guidance* reduces forecast error by 0.004 (0.277 × -0.013). Both are economically significant relative to the sample mean error of 0.076. By contrast, the coefficient of *GPT_Increm_Sentiment* is negative, suggesting that GPT's preference for more negative tone is associated with higher forecast errors. Crowley and Wong (2022) find that while negative language tends to be more salient, it frequently adds noise rather than improving predictive accuracy. This implies that GPT may give disproportionate weight to highly salient negative cues, even when doing so undermines performance. We find no association between incremental order/ length and GPT's forecast error.

GPT's quantitative reasoning, captured by *MetricOverlap*, is strongly associated with forecast accuracy. The coefficient of -0.015 implies that a one-standard-deviation increase in *MetricOverlap* corresponds to a 0.013 decrease in forecast error, or roughly 16.8 percent of the sample mean. The positive coefficient on *Post* (0.010, t = 3.10) is consistent with reduced forecast accuracy after the knowledge cutoff, in line with the view that information staleness is linked to weaker GPT performance (Levy, 2025).

Column (2) adds general press release characteristics as controls. The main findings remain unchanged, except that incremental guidance loses significance, suggesting that the positive effect observed in Column (1) was partly driven by the overall guidance content of the release. Among the controls, longer press releases in total words (*Log(PR_Total_Length)*), richer guidance content, and a more positive tone are all associated with lower forecast errors, whereas more obscure language and longer average sentence length (*Log(PR_General_Length)*) are related to greater errors.

Column (3) further includes firm size and analyst coverage. The coefficients on incremental readability, incremental sentiment, and metric overlap remain statistically significant, though the magnitude of the metric overlap effect falls to -0.005 ($p < 0.05$), still equivalent to 5.6 percent of the sample mean ($-0.005 \times 0.852/0.076$).[15] In line with the univariate evidence in Table 3, larger firms and firms with broader analyst coverage continue to exhibit significantly lower forecast errors (-0.018 and -0.024, respectively; both $p < 0.01$), consistent with the view that richer information environments and domain-learning opportunities support better GPT performance.

Taken together, the evidence in Table 4 suggests that GPT's forecast accuracy is heterogeneous and systematically related to its information processing features. However, a concern arises that some of the relationships we attribute to GPT's processing may reflect unobserved factors, such as disclosure quality, operational complexity, or macroeconomic conditions, that influence all forecasters rather than GPT alone. To address this concern and place GPT's performance in context, we next benchmark GPT forecasts against those of sell-side analysts, who are subject to the same underlying firm- and macro-level factors, and test

[15] The attenuation of the metric overlap result may reflect that it captures only one dimension of GPT's quantitative reasoning, whereas firm size and analyst coverage proxy for broader informational conditions that facilitate effective processing. We interpret metric overlap as an initial but incomplete window into GPT's internal reasoning, leaving scope for future work to identify additional dimensions.

whether the link between GPT's information processing and forecast accuracy continues to hold.

### *4.3 GPT's Forecast Accuracy Relative to Analysts*

We re-estimate Equation (3) using the difference in absolute forecast error, $|FE_GPT_{i,t}|$-$|FE_Analyst_{i,t}|$, as the dependent variable. Positive values indicate GPT underperforms analysts, while negative values indicate the opposite. This specification benchmarks GPT's forecasts against those of analysts for the same earnings announcement, helping account for unobserved factors that affect both forecasters. The goal of this comparison is not to evaluate GPT's relative advantage but to test whether its own information-processing features remain associated with forecast accuracy after accounting for these common influences.

The key predictors remain the incremental narrative features of GPT's selected sentences and the quantitative alignment indicator. If these variables capture GPT's own processing choices, we would expect their coefficients to retain both sign and significance in this relative specification. Conversely, if their effects in earlier tests merely reflect shared information conditions, they should attenuate toward zero once analyst performance is subtracted out.

Table 5 shows that the relationships identified in Table 4 remain robust when benchmarked against analyst forecasts. GPT's selection of more readable sentences and its high alignment with analysts' key metrics continue to be associated with smaller forecast error gaps, whereas a greater emphasis on negatively toned sentences remains associated with wider gaps. The magnitudes are smaller than in Table 4, consistent with some of these factors also influencing analyst performance. Overall, these results reinforce our main inference that GPT's forecast accuracy is systematically related to its information processing in both absolute and relative terms.

### *4.4 Analysis of GPT's Confidence Score*

In addition to accuracy, we examine GPT's expressed certainty about its forecasts. After generating each EPS prediction, GPT reports a continuous confidence score between 0 and 1 (*GPT_Confidence_Score*), with 1 indicating maximum confidence. This self-assessment provides a window into GPT's internal processing and allows us to test whether its expressed confidence corresponds to conditions under which it tends to perform well. A well-calibrated forecaster should express higher confidence when its processing choices and the surrounding information environment favor accuracy, and lower confidence when these conditions imply greater uncertainty.

For our sample, the mean (median) of GPT's self-assessed confidence score is 0.685 (0.700). Table 6 shows that GPT's confidence is correlated with several predictors of its forecast accuracy, including readability, sentiment, and guidance-related content in GPT-selected sentences, and quantitative alignment (*MetricOverlap*). These parallels suggest that GPT's self-assessment is at least partly grounded in features that are associated with stronger forecasting performance. However, confidence also increases with sentence length, even though it is unrelated to accuracy, and it remains elevated even when forecasts extend beyond the model's knowledge cutoff. These patterns suggest imperfect calibration: GPT's confidence appears to track cues associated with accuracy but does not consistently decline when its informational constraints are more binding, particularly in the post-cutoff period. This is consistent with Chen, Didisheim, and Somoza (2024), who document similar miscalibration in self-reported certainty scores. Nonetheless, the observed correlations indicate that GPT's expressed confidence carries informational value: while not a measure of accuracy per se, it provides additional insight into when the model's forecasts are likely to perform well or poorly.

### *4.5 Supplemental Analysis and Robustness Checks*

We conduct a series of supplemental tests to evaluate whether our inferences are robust to (1) alternative earnings definitions and horizons, (2) different model versions, and (3) more sophisticated prompts.

### *4.5.1 Alternative Forecast Targets and Horizons*

Our main analysis focuses on GPT's GAAP EPS forecasts, consistent with the accounting definition that is used in financial statements and increasingly adopted by analysts (Bradshaw et al., 2018). This provides a direct and conceptually aligned comparison between GPT's forecasts and reported financial results. Nonetheless, both academic research and professional practice also place considerable weight on alternative earnings measures, particularly non-GAAP EPS, which adjusts for one-time or non-operating items to highlight core operating performance. Because these adjustments are common in analyst valuation work, examining non-GAAP forecasts provides an important robustness check.

We prompt GPT to produce non-GAAP EPS forecasts and re-estimate our models. Internet Appendix Table IA4 reports results with non-GAAP EPS forecast errors as dependent variables. GPT successfully generates non-GAAP forecasts for 77.5 percent of sample firm-quarters (5,569 out of 7,114 firm quarters); missing cases arise when press releases lack explicit non-GAAP guidance. While GPT's adjustments do not always reconcile exactly with I/B/E/S street earnings, the non-GAAP results are largely consistent with those in the GAAP setting: GPT's forecast errors remain highly variable, and higher metric overlap and more favorable narrative selection are associated with greater accuracy. These findings suggest that the relationships between GPT's narrative and quantitative processing features and forecast accuracy extend across earnings definitions.

We also repeat the Table 4 analysis for one-quarter-ahead forecasts. Shorter-horizon forecasts reduce the influence of long-term uncertainty and are often considered easier to predict accurately than one-year-ahead forecasts. If GPT's performance is partly constrained by task difficulty, the one-quarter-ahead setting may provide a more attainable benchmark for the model. Internet Appendix Table IA5 shows that GPT's forecast accuracy improves at the one-quarter horizon, but the same relationships between processing patterns and forecast accuracy persist. This suggests that our main findings hold across forecast horizons and that variation in forecast accuracy reflects systematic processing patterns rather than task difficulty.

#### *4.5.2 Model Version and Prompt Format*

We use GPT-4 Turbo in our main analysis because its knowledge cutoff allows us to assemble data over a complete post-cutoff fiscal year—four quarters of data—for which realized one-year-ahead forecast errors can be calculated. A potential concern is that our findings reflect features unique to this specific model rather than general properties of large language models. To assess generalizability, we replicate our key analyses using OpenAI's GPT-4o, which offers enhanced multimodal capabilities but covers a shorter post-cutoff window. Internet Appendix Table IA6 shows that our main results remain consistent despite the smaller sample and higher risk of look-ahead bias.

We also assess the sensitivity of our results to prompt format. In particular, we design a chain-of-thought (CoT) prompt inspired by the CFA curriculum that mirrors the workflow of sell-side equity analysts conducting a bottom-up earnings forecast. This prompt incorporates explicit domain expectations (e.g., net-margin drivers, year-over-year growth), employs CFA-style terminology, and directs the model to reflect on, justify, and self-evaluate its assumptions. Internet Appendix Table IA7, Panel A reports the full prompt text, and Panel B shows that our two principal inferences remain unchanged: GPT's narrative selection and quantitative reasoning continue to be systematically associated with forecast accuracy. Overall, these findings suggest that the documented relationships between GPT's processing patterns and forecast accuracy are robust to model version or prompt format.

#### *4.5.3 Directional Forecasting Performance*

Our main analysis focuses on numerical forecast accuracy, as producing correct EPS figures represents the most stringent and information-rich test of forecasting ability.

Nonetheless, predicting the direction of future earnings changes can also be valuable, since even coarse directional calls may provide actionable trading signals. At the same time, directional accuracy discards information—both small and large signals map to the same binary outcome—making it a less precise measure of forecasting performance.

Using GPT's numerical GAAP EPS forecasts to infer the direction of earnings growth yields an accuracy rate of 49.3 percent (untabulated), close to a random coin toss. When prompted explicitly for direction, accuracy falls to 47.7 percent, well below the 71.1 percent achieved by analysts in our sample. While we attempt to replicate Kim et al.'s (2024) finding that LLMs outperform analysts on average in directional forecasts, our results do not provide supporting evidence, likely due to differences in sample, time frame, and task framing, as well as the look-ahead bias concerns highlighted by Levy (2025).

## 5 A Diagnostic Framework

The preceding analyses show that GPT's observable processing features—its narrative focus, quantitative reasoning, and self-assessed confidence—are systematically associated with variation in forecast accuracy. These features provide rare and measurable insights into the model's reasoning process. While users typically see only the final output, the intermediate steps—what GPT emphasizes and how confident it is—offer observable signals about the quality of the underlying analysis. Linking these processing features to forecast performance provides a structured approach to anticipating when model-generated information is likely to be more or less reliable.

Building on this insight, we propose a simple diagnostic framework that translates these observable processing features into a composite reliability score. Specifically, we construct an additive scoring system based on three dimensions:

1) Narrative attention (ranging from 0 to 3): one point each if GPT's selected sentences

are more readable, more forward-looking, and more positively toned than the average sentence in the same press release.

2) Quantitative reasoning (ranging from 0 to 3): one point for each financial metric identified by GPT that overlaps with those emphasized by sell-side analysts for the same firm and period, reflecting convergence with expert judgment.

3) Confidence (ranging from 0 to 1): one point if GPT's self-reported confidence score is above the sample median of 0.7. This indicator likely reflects aspects of model processing not captured by narrative attention or quantitative reasoning.

The resulting reliability score ranges from 0 to 7. The distribution is approximately bell-shaped, with 0.6, 3.9, 14.4, 23.7, 25.6, 20.2, 9.6, and 1.9 percent of forecasts receiving scores from zero to seven, respectively. Most observations cluster near the midpoint, and forecasts at the extremes are rare. This underscores the usefulness of diagnostic screening in identifying conditions that warrant greater caution.

Figure 2 plots average forecast error across reliability scores for the full sample (Panel A) and for subsamples partitioned by firm size and analyst following (Panels B and C, respectively). Due to the low frequency of score zero, we combine scores zero and one in the figures. In all cases, the relation is monotonic: forecasts with higher scores are consistently more accurate. The relation is economically meaningful. Panel A shows that GPT's forecast error declines from 0.13 at the lowest reliability scores (0, 1) to 0.03 at seven, consistent with GPT's forecasting performance improving at higher reliability levels and converging toward analyst accuracy. More importantly, we find a consistent pattern across subsamples, supporting the value of this diagnostic framework for firms operating in various information environments. It also suggests that the reliability score captures underlying variation in GPT's information-processing quality rather than firm-specific features or idiosyncratic noise.

This framework provides an interpretable and scalable approach to assessing when GPT's forecasts are more likely to be reliable. Conceptually, it transforms otherwise opaque model behavior into a set of observable, economically meaningful indicators—much like evaluating a chef's skill by observing their preparation rather than waiting solely for the final dish.

For researchers, this offers an empirical approach to capturing differences in model reasoning that relate to output quality and, more broadly, a means of quantifying a new dimension of information friction. For practitioners, it serves as a practical tool to screen model outputs and flag cases that may warrant additional scrutiny. For users, especially non-experts, it clarifies the sources of uncertainty in model output—uncertainty that, if unrecognized, may contribute to overreliance and widen informational disparities.

Although developed in the context of earnings forecasts, the same logic may extend to other domains where models interpret both narrative and quantitative disclosures, such as credit analysis, valuation, compliance monitoring, etc. Future research can test whether similar diagnostic signals predict reliability in these settings and explore how such mechanisms might help mitigate the challenges users face when evaluating model outputs.

## 6 Conclusion

Large language models such as GPT have the potential to democratize sophisticated financial analysis by lowering information-processing costs. But equal access does not necessarily lead to equal outcomes, as the locus of information frictions may shift from the cost of processing disclosures to the challenge of evaluating AI-generated outputs. This study examines how GPT processes financial disclosures and how observable aspects of that processing relate to output quality. In a realistic setting, i.e., forecasting one-year-ahead earnings from firms' earnings press releases, we analyze GPT's narrative attention and quantitative reasoning and link these processing patterns to variation in forecast accuracy, while refraining from causal claims.

Two main insights emerge. First, GPT's narrative attention is structured and consistent across firms, prioritizing sentences that are early, readable, forward-looking, and negatively toned. However, this consistency does not always translate into better accuracy: forecasts are more accurate when selected sentences are more readable and contain more guidance-related content but less accurate when GPT focuses on negatively toned sentences. Second, GPT's quantitative reasoning is context-dependent, improving when quantitative disclosures are clearer and when proxies suggest greater exposure during training. Forecast accuracy increases when GPT's selected metrics align with those emphasized by analysts, narrowing the gap with analyst benchmarks and challenging the view that LLMs perform uniformly poorly at numerical reasoning in financial contexts. GPT's self-assessed confidence is also positively associated with its processing quality.

Building on these results, we construct a diagnostic framework that aggregates three observable processing dimensions (i.e., narrative focus, quantitative alignment, and confidence) into a composite reliability score. The score provides a practical indicator of output reliability and captures a new form of information friction: the cost of evaluating model-generated information. While not a substitute for expert judgment, the resulting reliability score can alert users to contexts in which model-generated information merits greater caution.

Our findings suggest that advances in analytical technology may reduce traditional information-processing costs while introducing new evaluation frictions that influence how information is used in capital markets. Future research could extend this framework to analyze how these frictions affect price efficiency, investor learning, and disclosure design, and to examine whether similar diagnostic signals emerge in other domains such as credit risk assessment, valuation, compliance monitoring, etc.

**Declaration of Generative AI and AI-Assisted Technologies in the Manuscript Preparation Process**

During the preparation of this work, the authors used OpenAI's GPT models for two limited purposes: (1) generating research data through model interactions as part of the study's experimental design, and (2) assisting with minor language copyediting to improve clarity and readability. After using these tools, the authors reviewed and edited the content as needed and take full responsibility for the content of the article.

**Appendix A. Variable Definition**

| Variable Name | Definition |
|---|---|
| **GPT's Narrative Attention:** | |
| Picked_By_GPT | An indicator variable that equals 1 if a sentence is picked by GPT as important and 0 otherwise. |
| Order | The percentage ranking of the order of a sentence in a press release. |
| Length | The number of words in a sentence. |
| Fog | Readability of a sentence, calculated as the Gunning fog index. |
| Sentiment | The number of positive words minus the number of negative words in a sentence, divided by the total number of words in that sentence. Positive and negative words are defined by following the Loughran and McDonald sentiment word list. |
| Guidance | An indicator variable that equals 1 if the sentence contains any forecast and 0 otherwise. |
| | |
| **GPT's Quantitative Reasoning:** | |
| MetricOverlap | The number of overlapping financial metrics emphasized both by the GPT and analysts in IBES. We exclude the EPS and GPS metrics in IBES. |
| FS_LineItems | The total number of non-zero line items in the standardized quarterly and annual balance sheet and income statement that are fed to the GPT in our prompt. |
| Benf_Law_Deviation | Benford's law describes the relative frequency distribution for leading digits of numbers in datasets. For all non-zero numbers in the standardized balance sheet and income statement that are fed to the GPT in our prompt for each firm-quarter, we compare their leading digits' relative frequency to Benford's law and calculate the average deviation for digits between 1 and 9. |
| | |
| **Forecast Errors:** | |
| \|FE_GPT\| | GPT's absolute numerical GAAP-based EPS forecast error for the next fiscal year, calculated as \|GPT's one-year-ahead GAAP EPS Prediction - Actual IBES GAAP EPS value\|/Stock Price. |
| \|FE_Analyst\| | Analyst's absolute numerical GAAP-based EPS forecast error for the next fiscal year, calculated as \|median of Analyst's one-year-ahead GAAP EPS Prediction (GPS) - Actual IBES GAAP EPS value\|/Stock Price. |
| \|FE_GPT\|-\|FE_Analyst\| | The difference between GPT's absolute numerical GAAP EPS forecast error and analyst's absolute numerical GAAP EPS forecast error. |
| GPT_Confidence_Score | A GPT-generated continuous score that takes value from 0 to 1, indicating the level of certainty toward its own GAAP EPS prediction. |
| **Press-Release Characteristics:** | |
| PR_Total_Length | The total number of words in a press release. |
| PR_General_Length | The average number of words in all the sentences in a press release. |
| PR_General_Fog | The average fog index of all the sentences in a press release. |
| PR_General_Sentiment | The average Loughran-McDonald sentiment in all sentences in a press release. |
| PR_General_Guidance | The average of the “Guidance” dummy variable of all sentences in a press release. |
| GPT_Increm_Order | The average of the “Order” variable of the sentences picked by the GPT, minus the same average of all sentences in a press release. |
| GPT_Increm_Log(Length) | The average of the “Length” variable of the sentences picked by the GPT in log transferred form, minus the same average of all sentences in a press release in log transferred form. |
| GPT_Increm_Fog | The average of the “Fog” variable of the sentences picked by the GPT, minus the same average of all sentences in a press release. |
| GPT_Increm_Sentiment | The average of the “Sentiment” variable of the sentences picked by the GPT, minus the same average of all sentences in a press release. |
| GPT_Increm_Guidance | The average of the “Guidance” variable of the sentences picked by the GPT, minus the same average of all sentences in a press release. |
| **Knowledge Cutoff:** | |
| Post | An indicator variable that equals 1 if the current quarter earnings press release is announced after the knowledge cutoff date (4/30/2023) of the GPT-4 Turbo model and 0 otherwise. |
| **Firm Information Environment Proxies:** | |
| Firm_Size | Firm market value, calculated as the stock price times the number of common shares outstanding. |
| Analyst | Number of analysts following the firm. |

## Figure 1. Word Cloud Depicting GPT's Rationale

*Panel A. GPT's Rationale for Its Narrative Analysis*

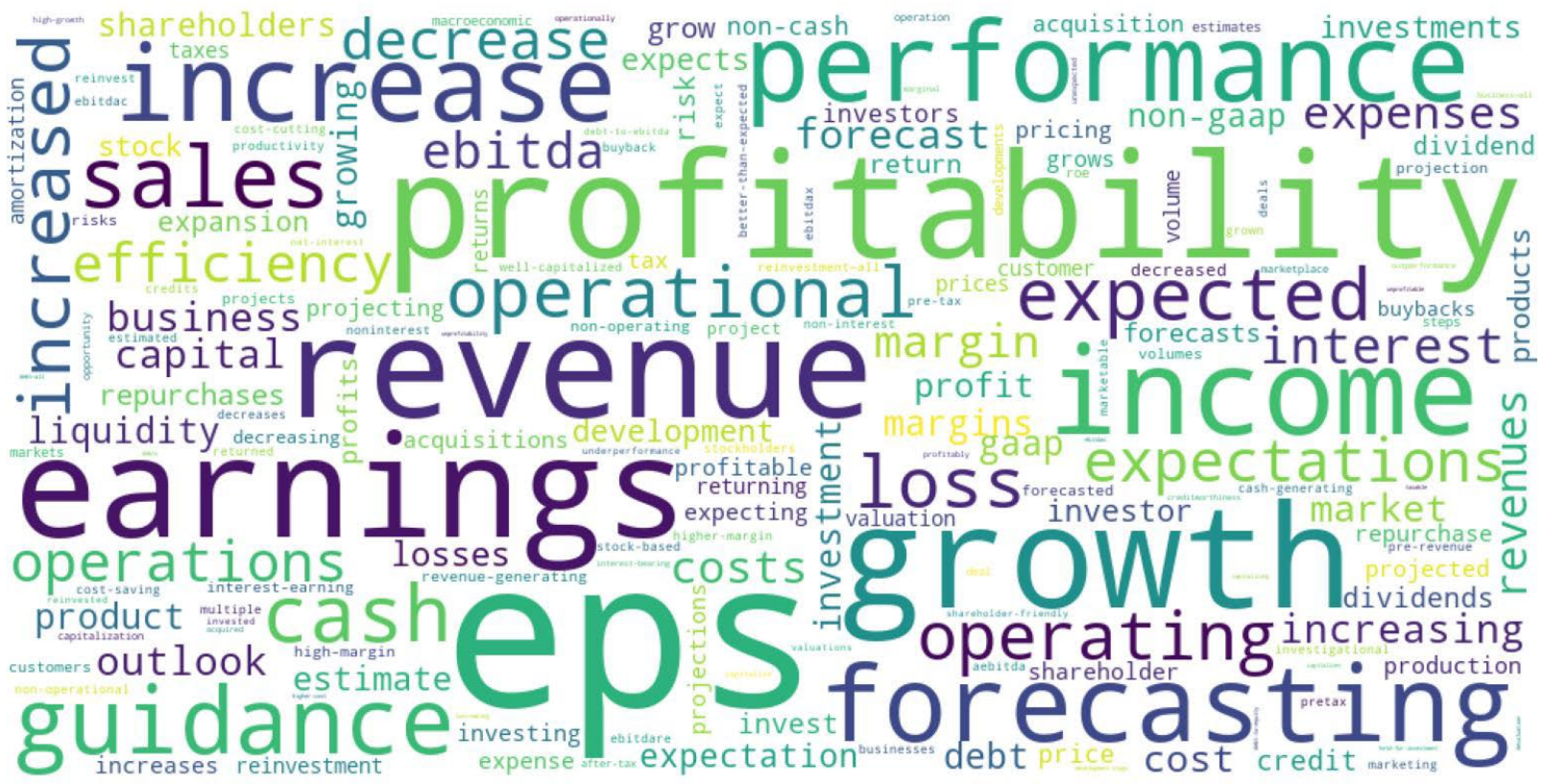

*Panel B. GPT's Rationale for Its Quantitative Reasoning*

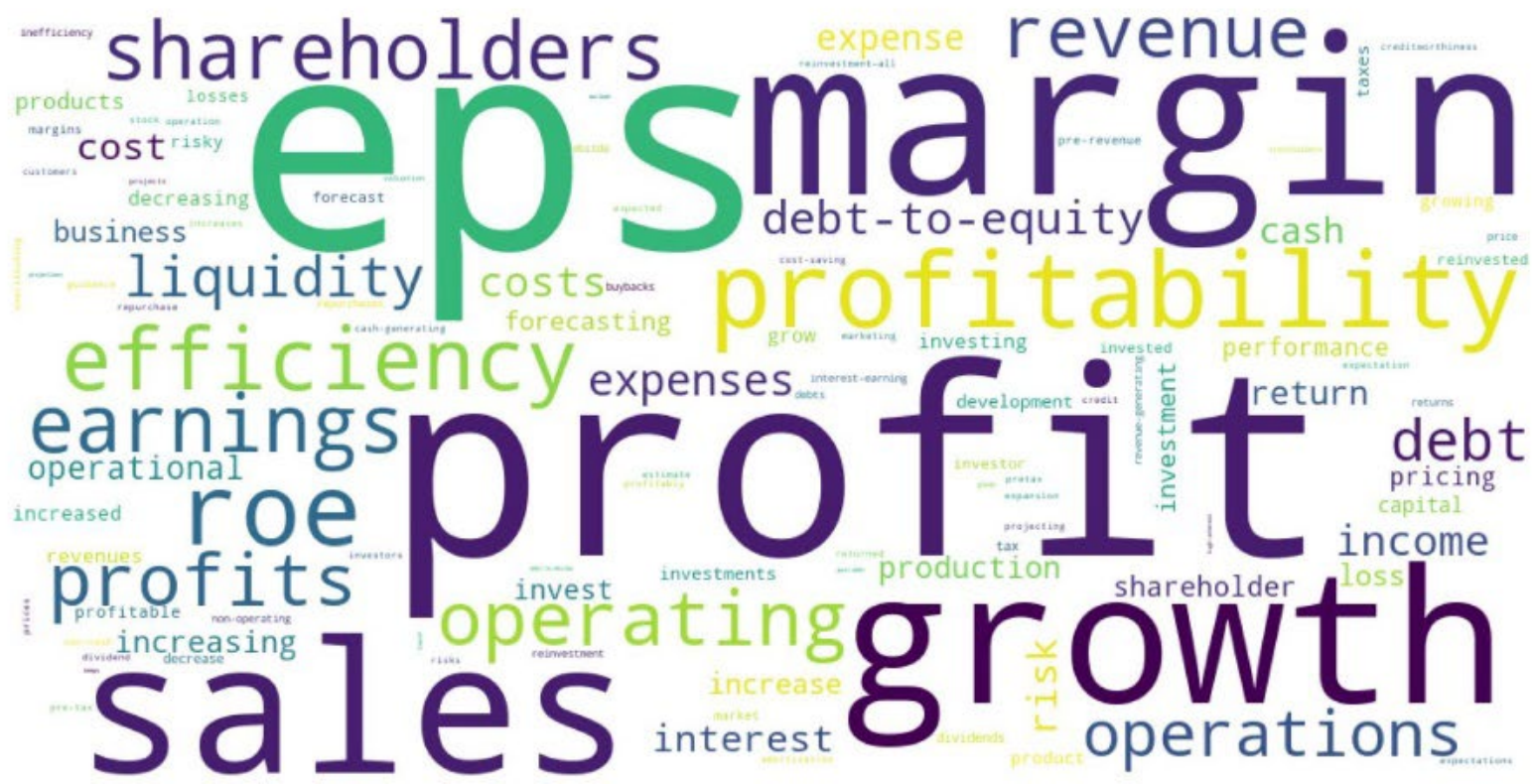

This figure presents word clouds from GPT-provided rationales. Panel A presents the most frequent words in GPT-provided rationales for its narrative analysis. In our prompt, we request GPT to select three sentences from a press release that it deems most important for EPS forecasting and explain why each of the selected sentences is relevant for its forecasting. Panel B presents the most frequent words in GPT-provided rationales for its quantitative reasoning. In our prompt, we request GPT to conduct quantitative analysis to forecast EPS and select three key financial metrics that it deems most important and explain why these selected metrics are crucial for this quantitative task. See Internet Appendix Table IA1 for complete details of our prompt to GPT.

**Figure 2. Relationship Between Reliability Score and GPT's Forecast Accuracy**

*Panel A. Average GPT's forecast errors |FE_GPT| of the full sample*

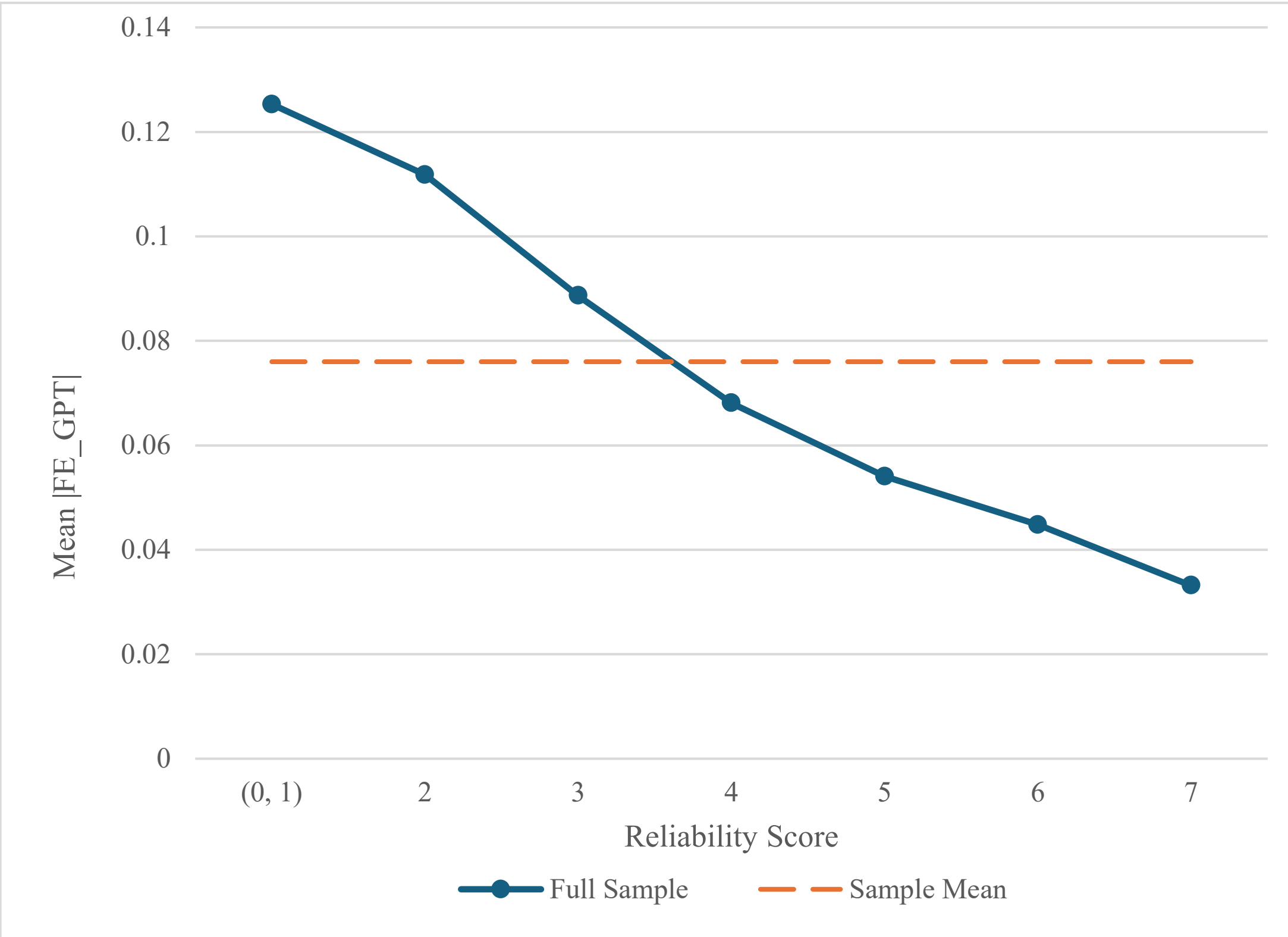

*Panel B. Average GPT's forecast errors by firm size*

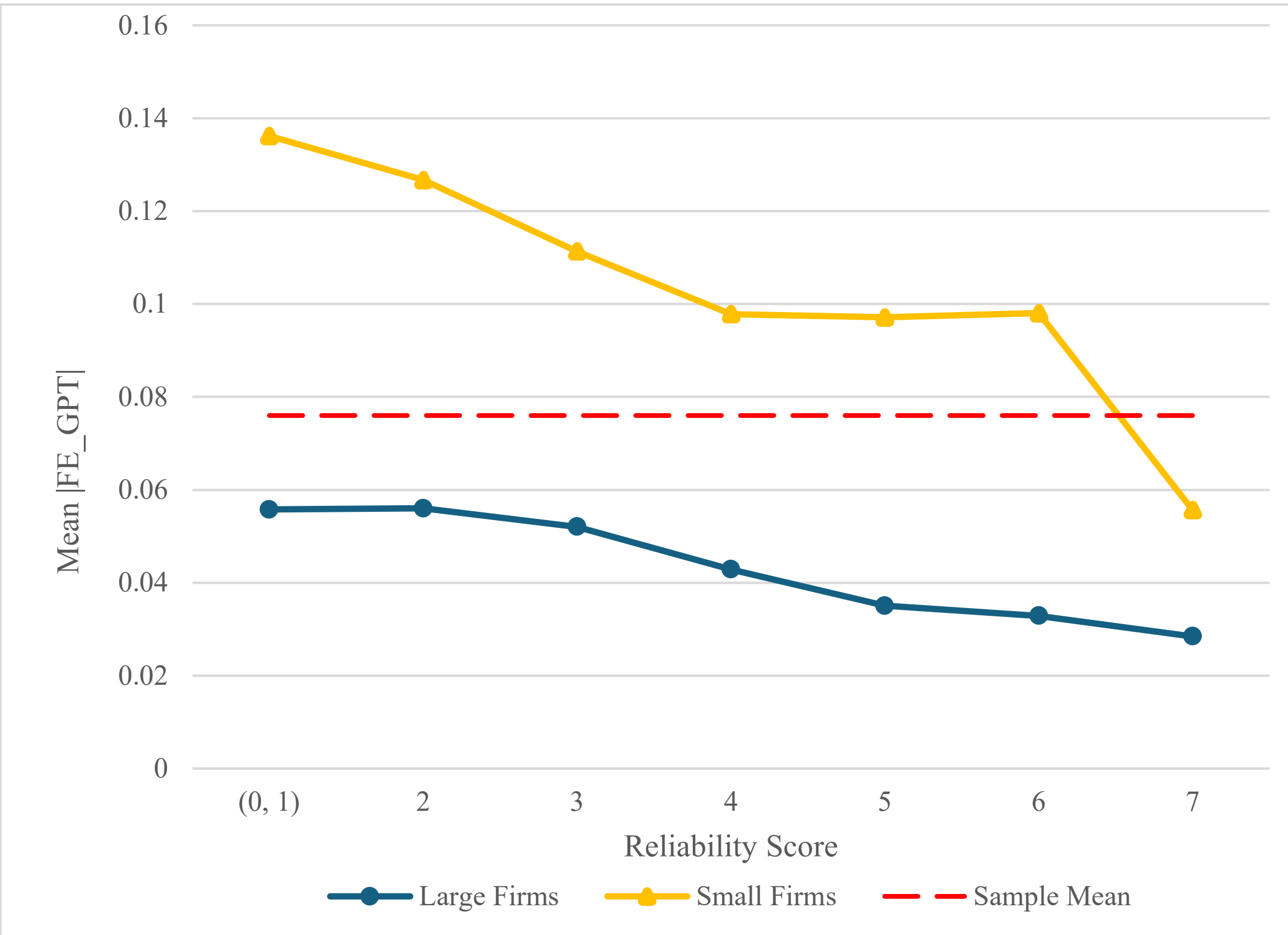

*Panel C. Average GPT's forecast errors by analyst following*

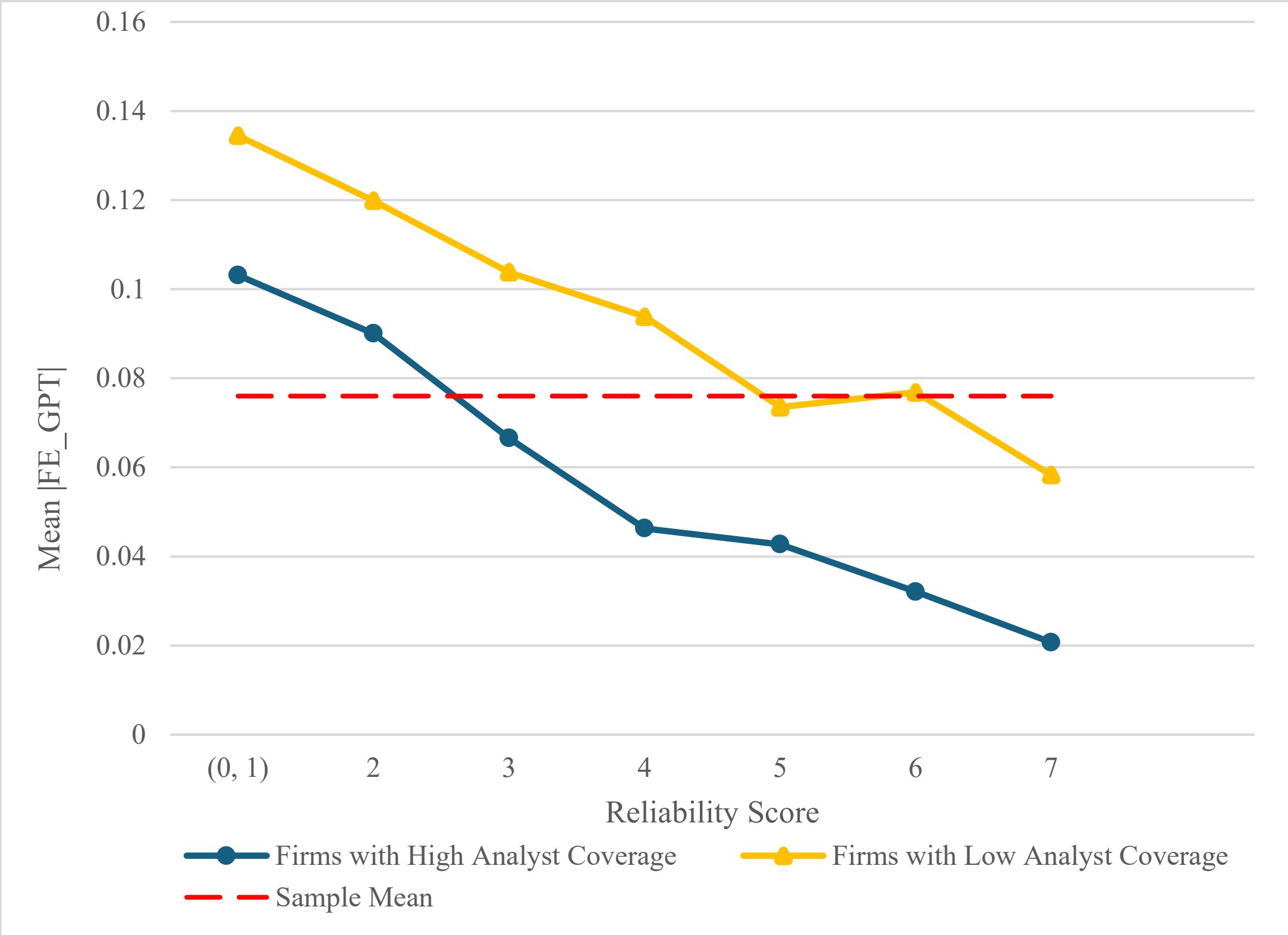

These graphs plot GPT's forecast errors across a composite reliability score. Panel A plots average GPT's forecast errors for the full sample, while Panels B and C present errors by subsamples partitioned by firm size and analyst following, respectively. Forecast error is as defined in Appendix A. The reliability score ranges from 0 to 7 and is constructed by aggregating three sets of observable processing features: (i) narrative attention (0–3), with one point each if GPT's selected sentences are more readable, more forward-looking, and more positively toned than the average sentence in the same press release; (ii) quantitative reasoning (0–3), with one point each for financial metrics identified by GPT that overlap with those emphasized by sell-side analysts for the same firm and period; and (iii) confidence (0–1), with one point if GPT's self-reported confidence score is above the median value of 0.7. Due to the low frequency of observations with a reliability score of zero, we combine scores zero and one in the graphs. The red dashed line in each panel represents the sample mean forecast error.

## Table 1. GPT's Narrative Attention

*Panel A. Summary statistics of sentence-level variables (329,333 sentences)*

| | Mean | Std. Dev | P25 | Median | P75 |
|---|---|---|---|---|---|
| Picked_By_GPT | 0.058 | 0.235 | 0.000 | 0.000 | 0.000 |
| Order | 0.511 | 0.289 | 0.261 | 0.511 | 0.761 |
| Length | 31.493 | 26.814 | 18.000 | 25.000 | 35.000 |
| Fog | 13.619 | 6.889 | 8.130 | 12.810 | 18.130 |
| Sentiment | 0.003 | 0.035 | 0.000 | 0.000 | 0.000 |
| Guidance | 0.054 | 0.225 | 0.000 | 0.000 | 0.000 |

*Panel B. LPM regression results of determinants of sentence selected by GPT*

| | (1) | (2) | (3) | (4) | (5) | (6) | (7) |
|---|---|---|---|---|---|---|---|
| | Full Sample | Firm Size | | Analyst Following | | Knowledge Cutoff | |
| | | Small | Large | Low | High | Pre | Post |
| Order | -0.110*** | -0.111*** | -0.109*** | -0.115*** | -0.103*** | -0.110*** | -0.111*** |
| | (-40.73) | (-31.62) | (-27.47) | (-34.76) | (-24.79) | (-36.89) | (-36.74) |
| Log(Length) | 0.088*** | 0.094*** | 0.082*** | 0.088*** | 0.089*** | 0.087*** | 0.089*** |
| | (36.04) | (28.20) | (24.67) | (28.68) | (23.50) | (33.49) | (33.93) |
| Fog | -0.005*** | -0.005*** | -0.006*** | -0.005*** | -0.006*** | -0.005*** | -0.005*** |
| | (-36.06) | (-26.04) | (-26.12) | (-26.52) | (-26.09) | (-33.01) | (-32.08) |
| Sentiment | -0.096*** | -0.145*** | -0.045** | -0.094*** | -0.101*** | -0.101*** | -0.091*** |
| | (-5.70) | (-5.81) | (-2.07) | (-3.91) | (-4.50) | (-5.00) | (-4.35) |
| Guidance | 0.182*** | 0.198*** | 0.170*** | 0.208*** | 0.158*** | 0.185*** | 0.178*** |
| | (29.97) | (21.62) | (22.07) | (25.40) | (19.82) | (27.20) | (27.09) |
| Press Release FE | Yes | Yes | Yes | Yes | Yes | Yes | Yes |
| N | 329,333 | 164,671 | 164,662 | 182,233 | 147,100 | 165,236 | 164,097 |
| adj. R-squared | 0.119 | 0.116 | 0.123 | 0.118 | 0.122 | 0.120 | 0.117 |

*, **, and *** indicate two-tailed statistical significance of 10%, 5%, and 1% levels, respectively.

This table reports analysis results of determinants of whether a press release is selected by the GPT as important (*Picked_by_GPT*) for 329,333 press release sentences. Panel A presents descriptive statistics, while Panel B presents Linear Probability Model (LPM) regression results for (1) the full sample, (2) and (3) the subsamples of small versus large firms based on the median value of firm size, (4) and (5) the subsamples of low versus high analyst following firms based on the median number of analysts, and (6) and (7) the subsamples of press releases issued in the period before and after the GPT-4 Turbo model's knowledge cutoff date (April 30, 2024). In all regressions we control for press release fixed effects and report t-statistics based on firm-clustered standard errors in parentheses. Variable definitions and data sources are detailed in Appendix A.

## Table 2. GPT's Quantitative Reasoning

*Panel A. Summary statistics of quantitative analysis variables (7,114 firm-quarters)*

| | Mean | Std. Dev | P25 | Median | P75 |
|---|---|---|---|---|---|
| MetricOverlap | 1.961 | 0.852 | 1.000 | 2.000 | 3.000 |
| FS_LineItems | 182.461 | 16.034 | 174.000 | 185.000 | 194.000 |
| Benf_Law_Deviation | 0.367 | 0.096 | 0.298 | 0.353 | 0.421 |
| Firm_Size | 13.163 | 38.452 | 0.655 | 2.220 | 7.207 |
| Analyst | 8.465 | 6.577 | 4.000 | 6.000 | 11.000 |

*Panel B. Regression results of GPT-analyst quantitative metrics alignment (7,114 firm-quarters)*

| | (1) | (2) |
|---|---|---|
| Log(FS_LineItems) | 2.111*** | 2.112*** |
| | (7.11) | (7.11) |
| Benf_Law_Deviation | -0.983*** | -0.983*** |
| | (-4.92) | (-4.91) |
| Log(Firm_Size) | 0.156*** | 0.156*** |
| | (8.53) | (8.53) |
| Log(Analyst) | 0.076** | 0.077** |
| | (2.08) | (2.08) |
| Post | | -0.004 |
| | | (-0.29) |
| Industry Fixed Effects | Yes | Yes |
| Fiscal Quarter Fixed Effects | Yes | Yes |
| adj. R-squared | 0.326 | 0.326 |

*, **, and *** indicate two-tailed statistical significance of 10%, 5%, and 1% levels, respectively.

This table reports analysis results of determinants of GPT-Analyst quantitative metrics alignment (*MetricOverlap*) for 7,114 firm-quarters. Panel A presents the descriptive statistics, while Panel B presents the OLS regression results. In all regressions we control for Fama-French 12 industry and fiscal quarter fixed effects and report t-statistics based on firm-clustered standard errors in parentheses. Variable definitions and data sources are detailed in Appendix A.

**Table 3. Summary Statistics of Forecast Errors**

*Panel A. GPT's forecast error |FE_GPT|*

| | N | Mean | Std. Dev | P25 | Median | P75 |
|---|---|---|---|---|---|---|
| \|FE_GPT\| | 7,114 | 0.076 | 0.141 | 0.007 | 0.025 | 0.075 |
| Pre-Cutoff | 3,530 | 0.071 | 0.127 | 0.008 | 0.025 | 0.073 |
| Post-Cutoff | 3,584 | 0.081 | 0.154 | 0.007 | 0.024 | 0.076 |

*Panel B. Mean of GPT's forecast error |FE_GPT| by rank scores of MetricOverlap*

| | N | Full Sample | Firm Size | | Analyst | |
|---|---|---|---|---|---|---|
| MetricOverlap | | | Small | Large | Low | High |
| 0 | 282 | 0.117 | 0.126 | 0.050 | 0.119 | 0.111 |
| 1 | 1,881 | 0.101 | 0.114 | 0.054 | 0.107 | 0.083 |
| 2 | 2,785 | 0.067 | 0.106 | 0.037 | 0.097 | 0.043 |
| 3 | 2,166 | 0.061 | 0.109 | 0.043 | 0.089 | 0.047 |

*Panel C. GPT's earnings forecast error relative to analysts |FE_GPT|-|FE_Analyst|*

| | N | Mean | Std. Dev | P25 | Median | P75 |
|---|---|---|---|---|---|---|
| \|FE_GPT\|-\|FE_Analyst\| | 7,114 | 0.035 | 0.099 | -0.001 | 0.007 | 0.038 |
| Pre-Cutoff | 3,530 | 0.033 | 0.095 | -0.001 | 0.007 | 0.036 |
| Post-Cutoff | 3,584 | 0.038 | 0.103 | 0.000 | 0.008 | 0.040 |

*Panel D. Mean of GPT's earnings forecast error relative to analysts |FE_GPT|-|FE_Analyst| by rank scores of MetricOverlap*

| | N | Full Sample | Firm Size | | Analyst | |
|---|---|---|---|---|---|---|
| MetricOverlap | | | Small | Large | Low | High |
| 0 | 282 | 0.063 | 0.067 | 0.038 | 0.056 | 0.080 |
| 1 | 1,881 | 0.046 | 0.051 | 0.030 | 0.046 | 0.046 |
| 2 | 2,785 | 0.031 | 0.047 | 0.018 | 0.045 | 0.020 |
| 3 | 2,166 | 0.028 | 0.055 | 0.018 | 0.045 | 0.020 |

Panel A reports GPT's earnings forecast error (|*FE_GPT*|) for the full sample and for pre- and post-knowledge-cutoff periods. Panel B reports the mean values of GPT's absolute earnings forecast error (|*FE_GPT*|) by *MetricOverlap* (rank scores 0–3) and further partitioned by firm size and analyst following. Panel C reports the difference in absolute forecast errors between GPT and analysts (|*FE_GPT*| - |*FE_Analyst*|) for the full sample and for pre- and post-knowledge-cutoff periods. Panel D reports the mean difference in absolute forecast errors between GPT and analysts by *MetricOverlap* and further partitioned by firm size and analyst following.

**Table 4. Analysis of GPT's GAAP Earnings Forecast Error**

*Panel A. Summary statistics of determinants of GPT's forecast error (7,114 firm-quarters)*

| | Mean | Median | Std. Dev. |
|---|---|---|---|
| **GPT-Selected Sentences' Incremental Characteristics** | | | |
| GPT_Increm_Order | -0.171 | -0.163 | 0.189 |
| GPT_Increm_Log(Length) | 0.528 | 0.509 | 0.581 |
| GPT_Increm_Fog | -2.148 | -2.608 | 4.302 |
| GPT_Increm_Guidance | 0.181 | 0.000 | 0.277 |
| GPT_Increm_Sentiment | -0.003 | -0.003 | 0.018 |
| **Quantitative Analysis Alignment:** | | | |
| MetricOverlap | 1.961 | 2.000 | 0.852 |
| **Knowledge Cutoff:** | | | |
| Post | 0.504 | 1.000 | 0.500 |
| **Press Release General Characteristics:** | | | |
| PR_Total_Length | 1505.032 | 1252.000 | 940.571 |
| PR_General_Length | 34.030 | 31.980 | 9.954 |
| PR_General_Fog | 13.729 | 13.646 | 2.681 |
| PR_General_Guidance | 0.064 | 0.045 | 0.071 |
| PR_General_Sentiment | 0.004 | 0.003 | 0.010 |
| **Information Environment Proxies:** | | | |
| Log(Firm_Size) | 1.501 | 1.169 | 1.291 |
| Log(Analyst) | 1.740 | 1.792 | 0.583 |

*(Continued on the next page)*

*(Table 4 continued)*
*Panel B. Regression results of GPT's earnings forecast error (7,114 firm-quarters)*

| | (1) | (2) | (3) |
|---|---|---|---|
| GPT_Increm_Order | 0.011 | 0.004 | -0.002 |
| | (1.18) | (0.43) | (-0.17) |
| GPT_Increm_Log(Length) | 0.002 | 0.001 | 0.001 |
| | (0.71) | (0.35) | (0.31) |
| GPT_Increm_Fog | 0.002*** | 0.003*** | 0.002*** |
| | (6.07) | (6.80) | (5.67) |
| GPT_Increm_Guidance | -0.013* | -0.002 | -0.006 |
| | (-1.92) | (-0.24) | (-0.87) |
| GPT_Increm_Sentiment | -0.561*** | -0.656*** | -0.623*** |
| | (-6.02) | (-7.04) | (-6.82) |
| MetricOverlap | -0.015*** | -0.013*** | -0.005** |
| | (-7.06) | (-6.16) | (-2.23) |
| Post | 0.010*** | 0.012*** | 0.012*** |
| | (3.10) | (3.60) | (3.72) |
| Log(PR_Total_Length) | | -0.016*** | -0.013*** |
| | | (-4.73) | (-3.94) |
| Log(PR_General_Length) | | 0.013* | 0.010 |
| | | (1.73) | (1.42) |
| PR_General_Fog | | 0.007*** | 0.007*** |
| | | (9.14) | (10.33) |
| PR_General_Guidance | | -0.104*** | -0.093*** |
| | | (-4.05) | (-3.70) |
| PR_General_Sentiment | | -1.624*** | -1.515*** |
| | | (-9.44) | (-8.98) |
| Log(Firm_Size) | | | -0.018*** |
| | | | (-10.52) |
| Log(Analyst) | | | -0.024*** |
| | | | (-7.46) |
| Industry Fixed Effects | Yes | Yes | Yes |
| Fiscal Quarter Fixed Effects | Yes | Yes | Yes |
| adj. R-squared | 0.053 | 0.078 | 0.113 |

*, **, and *** indicate two-tailed statistical significance of 10%, 5%, and 1% levels, respectively.

This table reports analysis results of GPT's one-year-ahead GAAP EPS forecast error (|*FE_GPT*|) for 7,114 firm-quarters. Panel A presents descriptive statistics of the determinant variables. Panel B reports OLS regression results of GPT's forecast error on the characteristics of GPT-selected sentences, GPT-Analyst quantitative metrics alignment, time relative to GPT's knowledge cutoff, and control variables. In all regressions we control for Fama-French 12 industry fixed effects and fiscal quarter fixed effects and report t-statistics based on firm-clustered standard errors in parentheses. Variable definitions and data sources are detailed in Appendix A.

**Table 5. Analysis of the Difference between GPT's and Analysts' GAAP Earnings Forecast Error (7,114 firm-quarters)**

| | (1) | (2) | (3) |
|---|---|---|---|
| GPT_Increm_Order | 0.002 | -0.001 | -0.004 |
| | (0.38) | (-0.23) | (-0.71) |
| GPT_Increm_Log(Length) | 0.002 | 0.001 | 0.001 |
| | (1.11) | (0.54) | (0.39) |
| GPT_Increm_Fog | 0.001*** | 0.001*** | 0.001*** |
| | (4.65) | (5.06) | (4.29) |
| GPT_Increm_Guidance | -0.008** | -0.002 | -0.003 |
| | (-2.10) | (-0.43) | (-0.74) |
| GPT_Increm_Sentiment | -0.225*** | -0.258*** | -0.243*** |
| | (-3.98) | (-4.54) | (-4.31) |
| MetricOverlap | -0.007*** | -0.006*** | -0.003** |
| | (-5.68) | (-4.80) | (-2.50) |
| Post | 0.003* | 0.004** | 0.004** |
| | (1.73) | (2.05) | (2.07) |
| Log(PR_Total_Length) | | -0.008*** | -0.007*** |
| | | (-3.76) | (-3.29) |
| Log(PR_General_Length) | | 0.012** | 0.011** |
| | | (2.55) | (2.42) |
| PR_General_Fog | | 0.003*** | 0.003*** |
| | | (5.92) | (6.50) |
| PR_General_Guidance | | -0.067*** | -0.063*** |
| | | (-4.27) | (-4.04) |
| PR_General_Sentiment | | -0.594*** | -0.559*** |
| | | (-5.66) | (-5.36) |
| Log(Firm_Size) | | | -0.008*** |
| | | | (-7.93) |
| Log(Analyst) | | | -0.005** |
| | | | (-2.31) |
| Industry Fixed Effects | Yes | Yes | Yes |
| Fiscal Quarter Fixed Effects | Yes | Yes | Yes |
| adj. R-squared | 0.049 | 0.061 | 0.074 |

*, **, and *** indicate two-tailed statistical significance of 10%, 5%, and 1% levels, respectively.

This table reports the OLS regression results of the difference between GPT's and analysts' one-year-ahead GAAP EPS forecast error (|*FE_GPT*|-|*FE_Analyst*|) on the characteristics of GPT-selected sentences, GPT-Analyst quantitative metrics alignment, time relative to GPT's knowledge cutoff, and control variables for 7,114 firm-quarters. In all regressions we control for Fama-French 12 industry fixed effects and fiscal quarter fixed effects and report t-statistics based on firm-clustered standard errors in parentheses. Variable definitions and data sources are detailed in Appendix A.

## Table 6. Analysis GPT's Confidence Level of Forecasting GAAP EPS (7,114 firm-quarters)

| | (1) | (2) | (3) |
|---|---|---|---|
| GPT_Increm_Order | -0.013** | -0.005 | -0.001 |
| | (-2.55) | (-0.96) | (-0.17) |
| GPT_Increm_Log(Length) | 0.004** | 0.003* | 0.004** |
| | (2.11) | (1.82) | (2.23) |
| GPT_Increm_Fog | -0.002*** | -0.002*** | -0.002*** |
| | (-8.59) | (-9.22) | (-8.18) |
| GPT_Increm_Guidance | 0.036*** | 0.023*** | 0.024*** |
| | (9.95) | (6.21) | (6.65) |
| GPT_Increm_Sentiment | 0.502*** | 0.610*** | 0.589*** |
| | (9.84) | (12.31) | (12.07) |
| MetricOverlap | 0.015*** | 0.012*** | 0.009*** |
| | (12.49) | (10.54) | (7.45) |
| Post | 0.001 | -0.001 | -0.001 |
| | (0.37) | (-0.53) | (-0.53) |
| Log(PR_Total_Length) | | 0.017*** | 0.016*** |
| | | (9.32) | (8.82) |
| Log(PR_General_Length) | | 0.001 | 0.001 |
| | | (0.21) | (0.36) |
| PR_General_Fog | | -0.005*** | -0.005*** |
| | | (-12.47) | (-13.35) |
| PR_General_Guidance | | 1.807*** | 1.769*** |
| | | (19.73) | (19.60) |
| PR_General_Sentiment | | 0.124*** | 0.119*** |
| | | (9.10) | (8.86) |
| Log(Firm_Size) | | | 0.012*** |
| | | | (13.36) |
| Log(Analyst) | | | 0.001 |
| | | | (0.41) |
| Industry Fixed Effects | Yes | Yes | Yes |
| Fiscal Quarter Fixed Effect | Yes | Yes | Yes |
| adj. R-squared | 0.127 | 0.194 | 0.219 |

*, **, and *** indicate two-tailed statistical significance of 10%, 5%, and 1% levels, respectively.

This table reports OLS regression results of GPT generated confidence score toward its EPS forecasts (*GPT_Confidence_Score*) on determinants for 7,114 firm-quarters. In all regressions we control for Fama-French 12 industry fixed effects and fiscal quarter fixed effects and report t-statistics based on firm-clustered standard errors in parentheses. Variable definitions and data sources are detailed in Appendix A.